**Enhancing Bridge Deck Delamination Detection Based on Aerial Thermography Through Grayscale Morphologic Reconstruction: A Case Study**

**Chongsheng Cheng**
The Durham School of Architectural Engineering and Construction
University of Nebraska-Lincoln
113 Nebraska Hall, Lincoln, NE 68588-0500
E-mail: cheng.chongsheng@huskers.unl.edu

**Zhexiong Shang**
The Durham School of Architectural Engineering and Construction
University of Nebraska-Lincoln
113 Nebraska Hall, Lincoln, NE 68588-0500
E-mail: szx0112@huskers.unl.edu

**Zhigang Shen, Corresponding Author**
The Durham School of Architectural Engineering and Construction
University of Nebraska-Lincoln
113 Nebraska Hall, Lincoln, NE 68588-0500
E-mail: shen@unl.edu

Word count: 4,332 words text + 11 tables/figures × 250 words (each) = **7,082** words

**ABSTRACT**

Environmental-induced temperature variations across bridge deck was one of the major factors that degraded the performance of delamination detection through thermography. The non-uniformly distributed thermal background yields the assumption of most conventional quantitative methods used in practice such as global thresholding and k-means clustering. This study proposed a pre-processing method to estimate the thermal background through iterative grayscale morphologic reconstruction based on a pre-selected temperature contrast. After the estimation of the background, the thermal feature of delamination was kept in the residual image. A UAV-based nondestructive survey was carried out on an in-service bridge for a case study and two delamination quantization methods (threshold-based and clustering-based) were applied on both raw and residual thermal image. Results were compared and evaluated based on the hammer sounding test on the same bridge. The performance of detectability was noticeably improved while direct implementation of post-processing on raw image exhibited over- and under-estimation of delamination. The selection of pre-defined temperature contrast and stopping criterion of iteration were discussed. The study concluded the usefulness of the proposed method for the case study and further evaluation and parameter tuning are expected to generalize the method and procedure.

**INTRODUCTION**

NDE method for detecting shallow delamination of bridge deck has been reported in variety researches (*1-4*). The principle of detection is based on the developed contrast in temperature induced by sun through the daily thermal cycle, so that the delaminated areas of bridge decks are exhibited as hotter or cooler regions in a thermal image at the heat-observing stage or heat-releasing stage correspondingly (*5*). Although the interpretation of the result is straight-forward, the quality of the thermal image is often degraded by factors such as the non-favorable time window of data collection, insufficiently developed temperature contrast, and surface inhomogeneities (*6-8*). Therefore, the engineering judgment of identifying delamination inevitably relied on expert's experience which was subjective to bias. To address the potential deficiency and improve the productivity of image processing, several quantitative methods have been developed to process the thermal image based on temperature contrast, temperature gradient, and temperature density distribution (*1-3; 9-11*). However, these methods suffered degraded performance under spatial temperature variations, which were often referred as the non-uniform thermal background. This issue was observed in both experimental setup when using a non-uniform excitation heating source, and under natural outdoor environment. The underlined assumption of these processing methods required a relative uniformed background to represent the "sound" concrete area so that the highest temperature areas were associated with the delamination. However, this requirement cannot always be satisfied (*12*). Thus, the delaminated area shown as regional maxima were often missed. To address the issue and improve the performance of the detectability, a pre-processing procedure is desired that could be used for accounting the non-uniformity of the background.

Image processing methods of delamination quantification for bridge deck are often contingent on the field of view for the image to be processed and the associated assumption. The field of view for the thermal imaging system is determined by the camera specification and data collection method. For rapid data collection, the thermal image could be obtained by the imaging system (e.g. IR camera) through different configurations: hand-held or tripod mounted (*1; 3*), ground vehicle mounted (*2*), or aerial vehicle mounted (*4; 10*). And for a single image, the field of view was determined by the lens angle and the mounting height when facing the camera downward. For the camera with the same lens, increase the mounting height, the field of view increases which gives larger scanning area included in the single image. Thus, the hand-held or car-mounted setups provide less field of view comparing to the UAV (unmanned aerial vehicle) mounted setup due to the different surveying height capability. The field of view could be increased by tilting the camera that the view axis was away from 90 degrees to the ground. The more tilting of the camera, the more perspective distortion brings to the scanning area which would require more correction for the pre-processing of data. After the acquisition of data, the detection method could apply on the single image or the stitched image covering entire bridge decks. The essential is depend on the content included in the field of view and what assumption was made based on the interpretation of the thermal image. For each image to be processed, it could contain three scenarios in terms of the field of view: sound area only, a delaminated area only, or both of sound and delaminated area. Dabous et al. (*3*), Kee et al. (*9*), Oh et al. (*1*), and Vaghefi et al. (*2*) converted the raw thermal image into binary image to locate the delamination through thresholding a specific temperature percentile or value on each image by assuming the high temperature shown in delaminated area. Due to the observed large temperature variation across the entire bridge deck, no single threshold could be used as the global criterion to distinguish the delaminated area from the sound area (*1; 3; 12*). Thus, the threshold method worked appropriately for the scenario containing both delaminated and sound areas while the other two scenarios required the operator's experience. Abdel-Qader et al. (*13*) developed an automatic process to segment the delaminated area based on the region-growth algorithm. The method assumed the delaminated area shared the highest temperature over the scene and thus used the neighbor temperature deviation difference inside a 9-by-9 pixel window as the criterion for screening the image. Ellenberg et al. (*10*) extended the method through using temperature gradient as the threshold criterion. Although the region growth-based segmentation method showed the robustness for all scenarios of the field of view, the assumption was still yielded by the delaminated area to be the global maxima in temperature (*13*) or gradient (*10*). Omar and Nehdi (*11*) developed a k-means based clustering model to detect the delamination of bridge deck by screening on the condition of deck age, temperature contrast, and the dimension of founded spalling over the bridge decks. Based on the criterion of the above condition, the number of K could be determined so that the thermal image could be classified into k groups without the direct interpretation of the temperature variation. Since the k-means clustering algorithm is based on the variation of density distribution, it is best for the field of view that covers whole bridge deck without large background variation. In all, the previous processing methods were shown the ability of automatic processing in different scenarios of the field of view but lacked the consideration to account for the background variation presented in the large area coverage (such as entire bridge) in the field of view.

## RESEARCH OBJECTIVES

The objective of this study is to develop a pre-process procedure, using grayscale morphology, to account for the non-uniform temperature variation presented in the thermal images. The estimation of background will be achieved through an iterative reconstruction of the thermal image through grayscale morphologic operations with the help of a pre-selected temperature contrast value. After the reconstruction of the background, the residue image could be subtracted from the original image for successive use. In this paper, a case study using the proposed approach was presented with improved results compared to existing threshold-based and clustering-based methods.

## BACKGROUND

### Temperature Contrast as The Pre-selected Criterion

Although the temperature contrast as the criterion for detecting delamination has been widely studied, the issue remains on the selection of a well-identified reference as the background. According to ASTM (D4788-03), at least 0.5 $^{\circ}$C difference between the delaminated area and adjacent sound area was recommended. This recommendation only describes the minimum temperature contrast under the ideal condition, but the contrast varies in practice in terms of different delamination size and depth (*7*) as well as the different time window of a day (*6; 9*). Recently, a comprehensive study was conducted by Hiasa, Birgul and Catbas (*14*) to evaluate the detectability of delamination detection by thermography in terms of different sizes, shapes, and depth throughout the finite element simulation. The study concluded that with the increase of delamination area, the temperature contrast increase until stable when the size getting large enough (e.g. over 40x40 cm). Also, it showed the temperature contrast reached positive maximum (the delaminated area was higher than sound area) during the noon to afternoon. Thus, it recommended the 0.4 $^{\circ}$C as the lower bond of the contrast for the certain detectability. Sultan and Washer (*12*) conducted a reliability analysis of delamination detection by thermography based on the receiver operating characteristics (ROC) of the temperature contrast. It evaluated the effect of spatial temperature variation on the contrast-based thresholding method and the optimum contrast was suggested 0.6 ºC for in-service bridge deck and 0.8 ºC for a simulated slab with a considerable accuracy (80%). Even though the effort has been made for the selection of the rational temperature contrast, the method of utilizing the contrast criterion on the background with spatial temperature variation does not yet cause fully attention. The challenge remained is the determination of non-uniform background without prior awareness of delamination allocations. Thus, this study raises the research question that can reconstruct the background (sound areas) from delaminated areas as the regional maxima with the help of the well-studied temperature contrast?

**Grayscale Morphologic Reconstruction for Background Estimation**
Grayscale morphology extended from mathematical morphology was a useful approach for image processing based on the geometric shape of the object presented in the image (*15*). It is based on the set theory and consisted of the combination of dilation, erosion, and set operations (e.g. union, intersection, and complement). It has been used for image filtering, edge detection, denoising, region filling, and segmentation (*16-18*). The applications in civil engineering have been found for concrete material analysis (*16*), region recognition from LiDAR data (*19; 20*), and pavement crack detection (*21; 22*). The advantage of using the morphologic operation was to transform the image based on the object's shape so that geometric information could be controlled throughout the manipulation. The basic operation of grayscale morphology consists of the dilation and erosion which exhibits the behavior of expanding and shrinking. The dilation ($\oplus$) of an image ($F$) is the maximum filtering process with a structural element ($B$) which is defined in Equation 1 while the erosion ($\ominus$) is the minimum filtering process in Equation 2 below.

$$F \oplus B = \max\{f(x,y) \mid x, y \in B\} \text{ , where B presents the structure } \text{✱} \text{ for 8-connectivity} \qquad (1)$$

$$F \ominus B = \min\{f(x,y) \mid x, y \in B\} \text{ , where B presents the structure } \text{✱} \text{ for 8-connectivity} \qquad (2)$$

Image is often presented in the form of a 2D matrix and Figure 1 illustrates the dilation and erosion processing on a matrix. The dilation process on a matrix transforms each entry to the maximum value based on its neighbor connectivity defined by the structure element and vice versa for the erosion process. Thus, the dilation of an image often returns an enlargement of object's bright shape and the erosion returns the shrinkage of the bright shape (see Figure 2, bright regions have the value of 1 and dark regions have the value of 0).

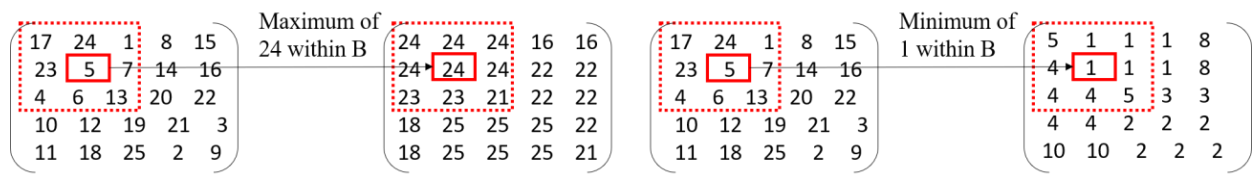

**FIGURE 1 Example of Dilation (left) and Erosion(right) of a matrix.**

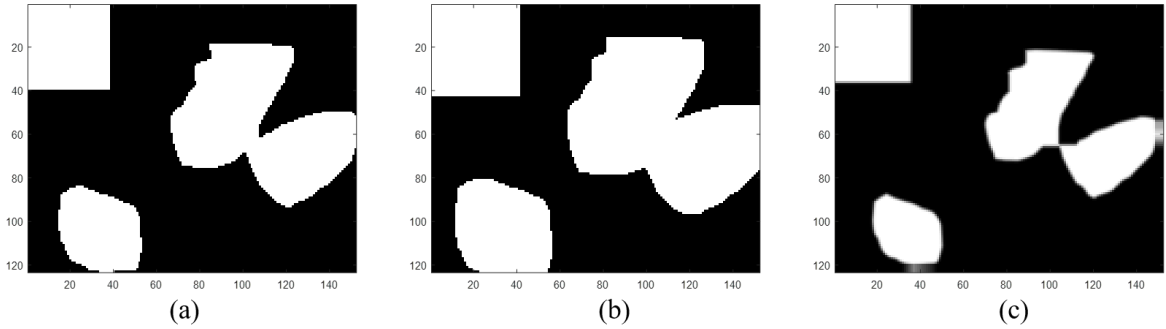

(a)　　　　　　　　　　　(b)　　　　　　　　　　　(c)

**FIGURE 2 Illustration of dilation and erosion on binary image: (a) original synthetic image; (b) dilate 3 times of (a); erode 3 times of (a)**

Grayscale morphologic reconstruction is an iterative operation based on dilation and erosion that often used for dome (regional maximal) and basin (regional minima) extraction. Given two functions to represent a mask image ($F$) and a marker image ($G$) which $G<F$ meaning $G$ "under" $F$. The reconstruction by dilation is then defined below:

$$R_F^D(G) = D_F^k(G), \text{ when } D_F^k(G) = D_F^{k+1}(G) \tag{3}$$

Which $D_F^1 = (G \oplus B) \wedge F, D_F^k = D_F^1(D_F^{k-1}(G))$ (4)

Where $\wedge$ is the inclusion operation from Set theory and a "pointwise minimization" to ensure the $(G \oplus B)$ is "under" the mask $F$. Thus, the reconstruction by dilation $R_F^D(G)$ can be interpreted as the marker image $G$ dilates with the structure element $B$ under the mask image $F$ until stability is reached ($D_F^k(G) = D_F^{k+1}(G)$). When the maker image $G = F - h$, a special case that marker image $G$ is the shifted $F$ by offsetting $h$, the reconstruction can be used for dome extraction. Figure 3 illustrated the 1D representation of this procedure. Based on the definition of reconstruction, the key to extract the regional maxima is depended on the selection of $h$ value which defines the contrast between maxima and its surroundings. This property becomes useful for our case since the temperature contrast is the key to distinguish delamination from sound area.

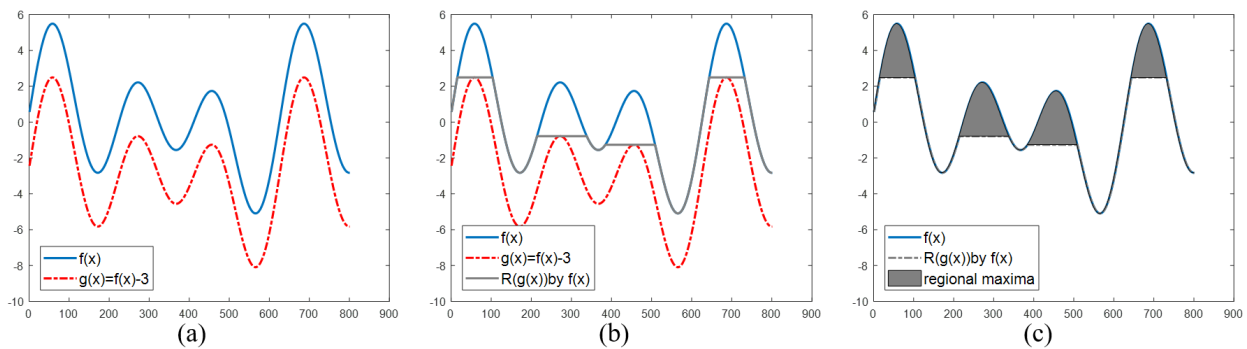

(a)　　　　　　　　　　　(b)　　　　　　　　　　　(c)

**FIGURE 3 Illustration of grayscale morphologic reconstruction for 1D representation: (a) synthetic 1 D signal f(x) = sin(x)+2\*cos(2\*x+5) +3\*sin(3\*x) and g(x) = f(x)-3; (b) reconstruction of g(x) by f(x); (c) regional maxima by difference between f(x) and R(g(x))**

**METHODOLOGY**

The thermal background is estimated through an iterative the grayscale morphologic reconstruction using the temperature contrast as the $h$ criterion. Figure 4 shows the procedure for thermal background estimation. The original thermal image (T) was shifted to a marker image (G) by subtracting a pre-defined temperature contrast ($h = 0.5$°C). Then the reconstructed image G' was calculated through grayscale morphologic reconstruction of G by T. The different was then calculated for the reconstructed image ($G'_n$) and the previous one ($G'_{n-1}$). If there would exist a maximum value in the difference smaller than $h$, it indicated the difference between two backgrounds was not identical under the criterion of $h$ so that it could be used as the final background. Otherwise, it would be resent to the reconstruction process until satisfying the criterion. After the background was estimated, the residual was then calculated by original image subtracted from the background. The residual would contain the regional maxima derived from the background.

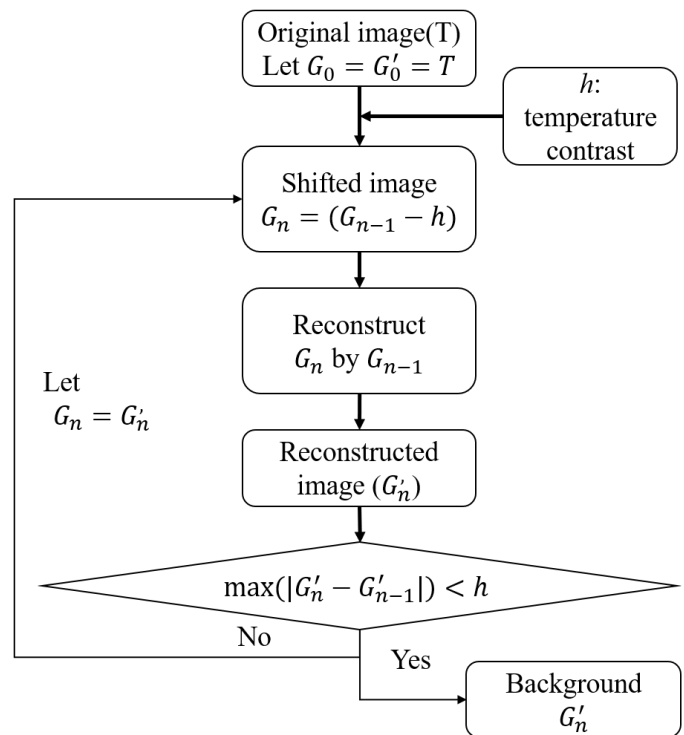

**FIGURE 4 Thermal background estimation**

**FINDINGS OF THE CASE STUDY**

A field study was conducted for delamination detection through aerial thermography. The survey was carried out for an in-service bridge at the northbound of US 77 close to Lincoln, Nebraska. The data was collected through UAV with carrying optical and infrared cameras in October 2017. Two configurations were implemented in Figure 5 based on different payload requirements. DJI matrix was used to carry the infrared camera integrated with an onboard computer for thermal imaging. DJI inspire one carried the High Definition visible camera for imaging the surface condition of the bridge. A pilot was responsible for controlling the UAV while the imaging system was set as automatic recording during flight. The path of UAV was designed to follow the center of the bridge at a fixed height and constant speed without traffic control. After the data was collected, the thermal image was then pre-processed through the customized MATLAB algorithms for perspective correction, rotation, and stitching. Then the accuracy of the stitched thermal image was checked and calibrated through manual comparison with the visible image for this case.

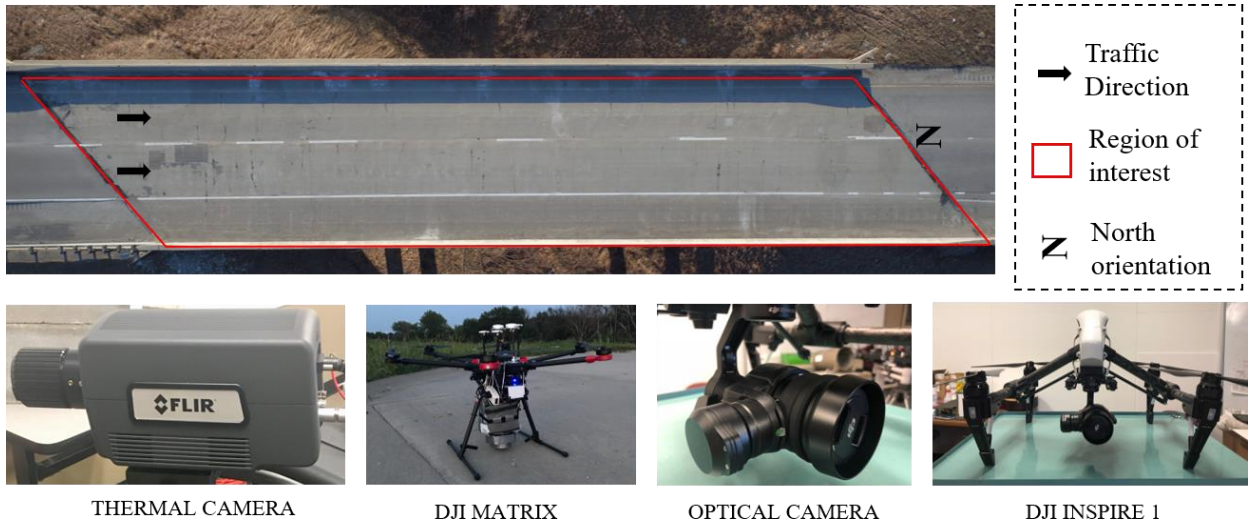

**FIGURE 5 Field setup and data collection**

Figure 6 shows the thermal raw data for the bridge. The temperature of the bridge was color-coded where high to low temperature was represent through red to blue. the temperature variation was observed across the bridge. The highest temperature was observed about 33 °C which was around at the joint over the abutment while the lowest was around 19 °C for the top shadow. Since the data was collected in afternoon, the potential delamination would be appealed as hot regions. The candidate the regions had a temperature variation from 30.7 °C to 27.3 °C. Besides the hot regions, the potential sound area also had a temperature variation from 27.5 °C to 26 °C. This phenomenon had been widely observed in previous studies which made the global threshold method performed unsatisfactorily so that it was more rational to focus on the regional maxima.

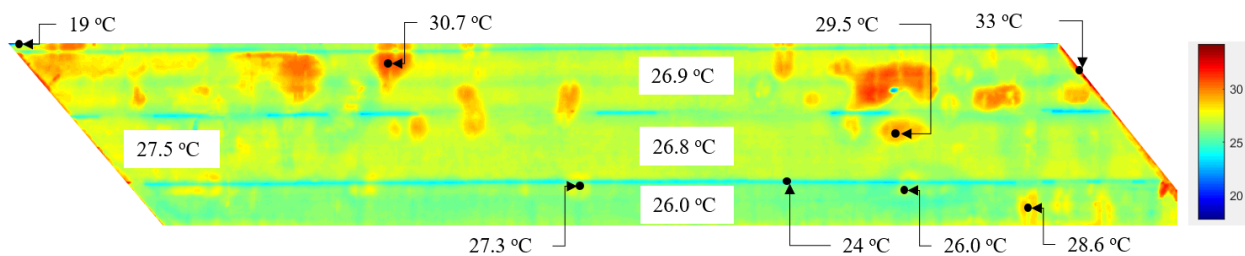

**FIGURE 6 Thermal raw data**

Figure 7 shows the intermediate results of background estimation were obtained during the iteration, and the hammer sounding result (validated by coring) was used as the reference for ground truth (Figure 9). In Figure 7, the regional maxima in the processed image were gradually "faded away" until 10th iteration which the difference between iteration 9 and 10 was not significant. Thus, the result from iteration 9 was used as the background and residual image was then calculated (see Figure 8). In Figure 8, the regional maxima were derived from the background, and the information of variation for maxima was kept so that the post-processing such as segmentation could be used for discrimination of delamination. Figure 9 shows the hammer sounding results provide by NDOT which the locations of delamination were marked and extracted. The spatial distribution of delamination in the hammer sounding map was very similar to the thermal image (Figure 6) and the residual image (Figure 8).

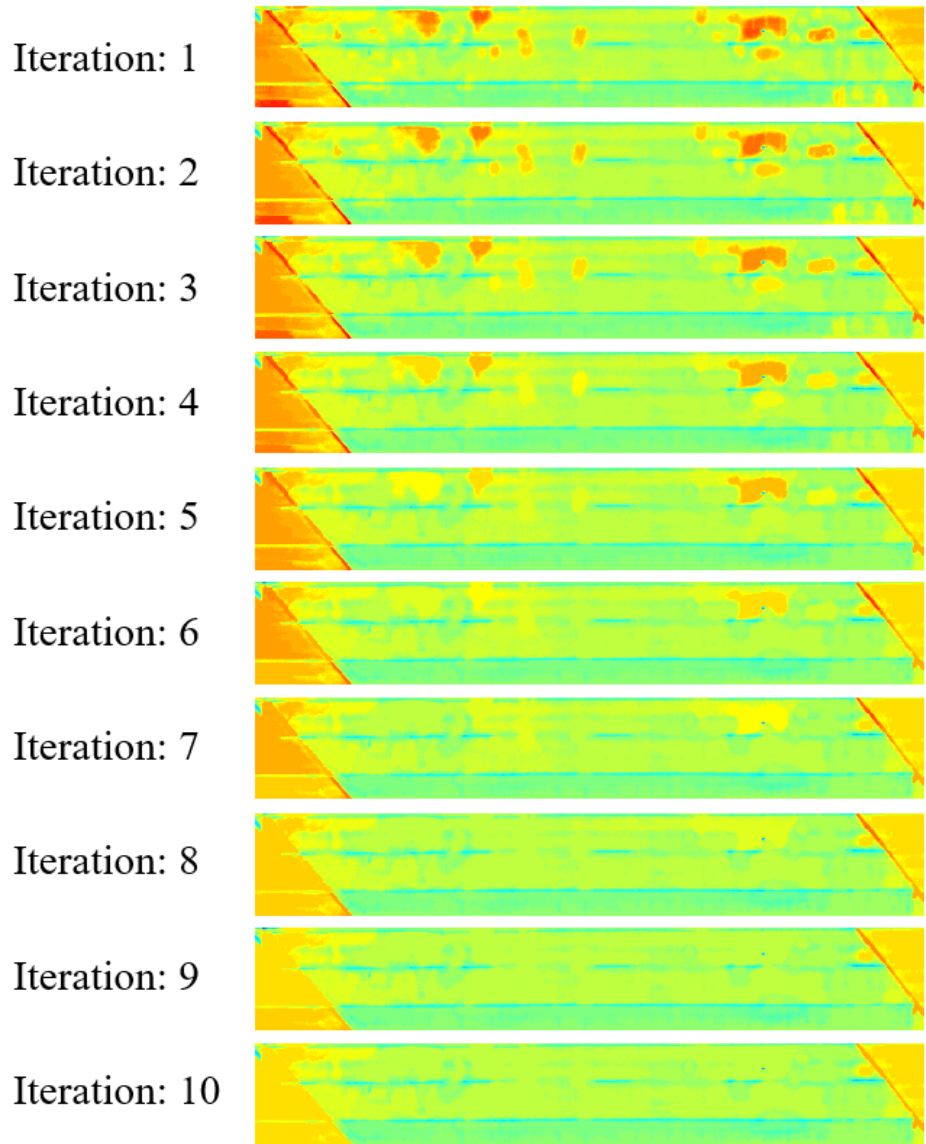

**FIGURE 7 iteration of background estimation**

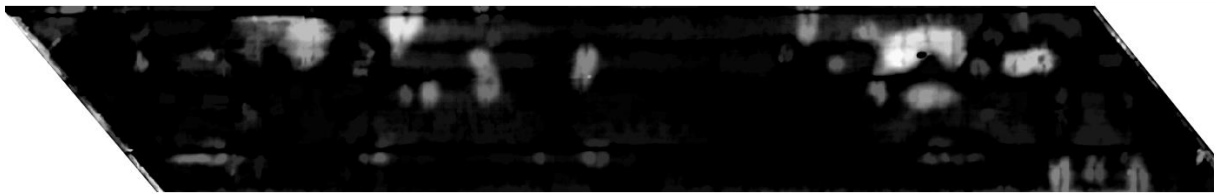

**FIGURE 8 Residual image representing regional maxima**

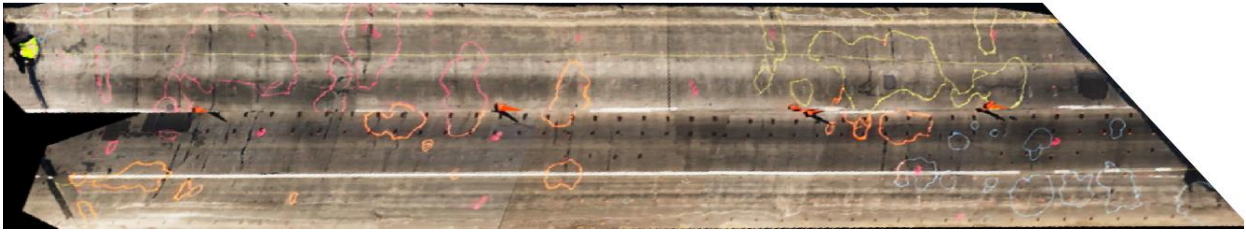

**FIGURE 9 hammer sounding result**

The processed results are compared between using raw image and using residual image. Two methods were tested: 1) global threshold; 2) K-means clustering. In Figure 10, a comparison was made to implement the threshold on the raw image and the residual image. Top image (Figure 10 top) shows an underestimation of delamination when the top-half image had a similar pattern to the hammer sounding result, the bottom-half suffered the insufficient estimation. Figure 10 bottom shows the overestimation when the bottom-half image had the sufficient indication of delamination, the top-half image shows overestimated. The residual image (Figure 10 middle) shows closer estimation in terms of the threshold-based method.

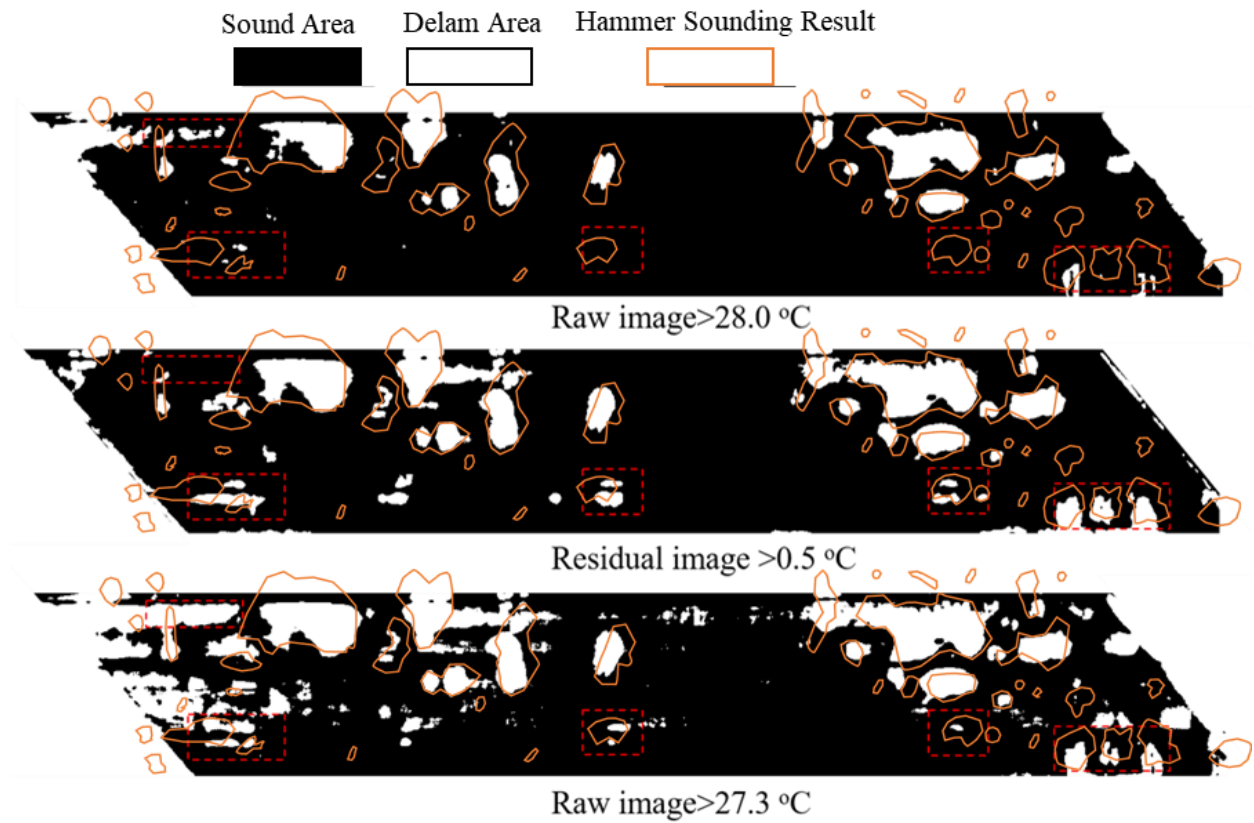

**FIGURE 10 Delamination detection based on global threshold method**

K-means clustering as a multiple level classification method was recently introduced by (*11*) for delamination indication. The selection of number K is often based on researcher's experience or follow the criterion proposed by (*11*). We compared the situation of k equals 2 and 3 on the raw and residual image. Three color codes (white, gray, and black) were used to indicate the conditions of delamination, possible delamination, and sound in Figure 11. When k equals 2, the k-means method showed similar outcomes for both raw and residual data. However, when k equals 3, the outcome behaved very differently, and the residual data showed closer prediction comparing to the raw data. The reason could be the k-means method miss-classified the intermediate level of background on the top-half image with the delamination on the bottom-half image due to the close temperature value shared.

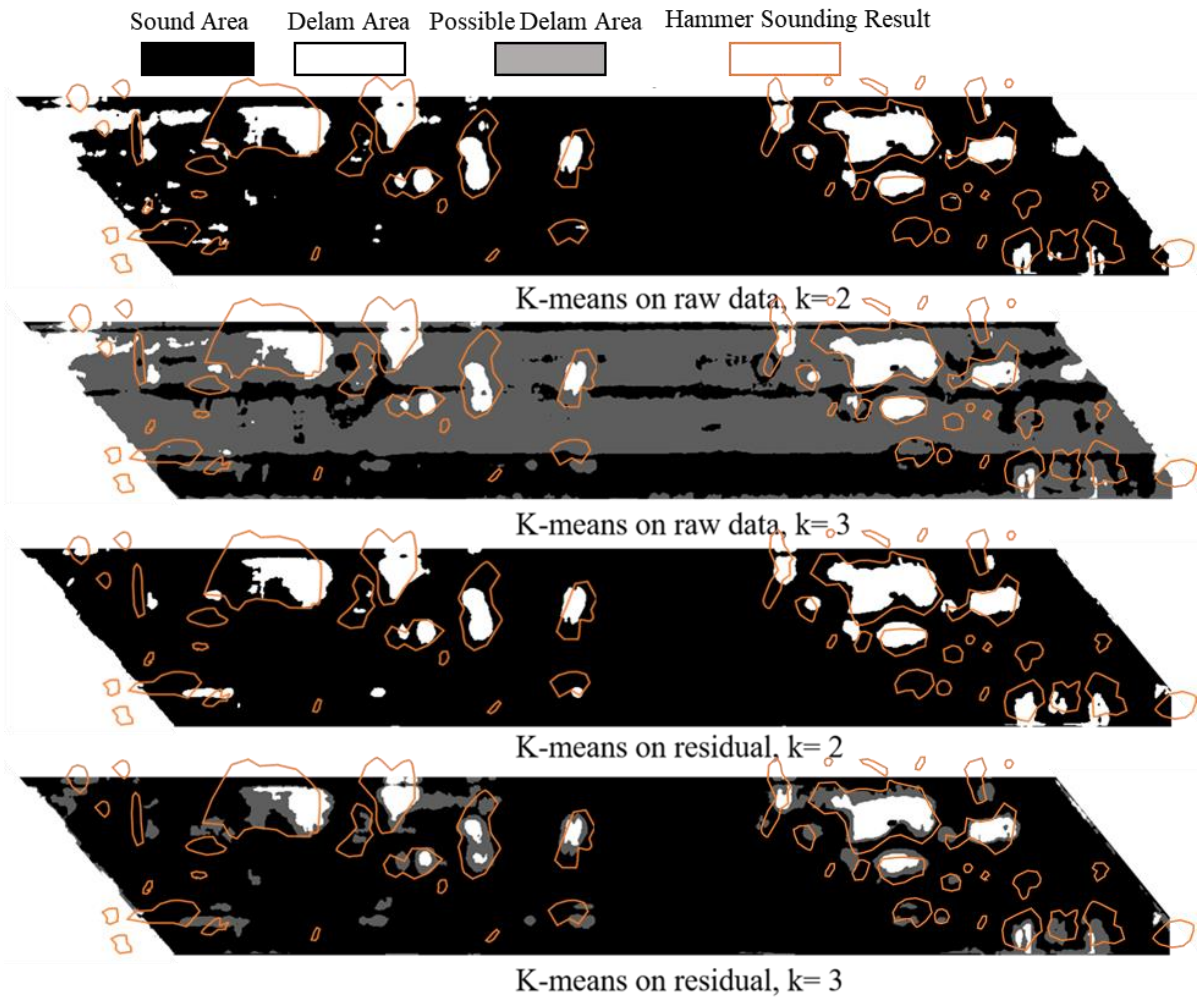

**FIGURE 11** Delamination detection based on clustering method

## DISCUSSION

The proposed method showed the capability to enhance the performance of detection for both threshold-based and clustering-based methods. Since neither threshold-based nor clustering-based method took account for the presence of temperature variation (non-uniform thermal background), this variation did cause the under- or over-estimation for threshold method and misclassification for clustering method (see Figure 10 and Figure 11 middle). The processed method as a pre-processing for background estimation, enabled the ability to remove the effect of temperature variation. However, we observed the pattern generated from the proposed method had similar shapes and locations, but with reduced size compared to the hammer sounding result. The underlined reason was expected by the diffusive nature of the temperature feature for delamination, and the contrast value selected in this case. Even though the non-uniform thermal background has been accounted, the thermal feature of delamination is still time and environment dependent and requires the sufficient experience during the data collection.

The shifted contrast *h* and stop criterion during iteration were two key parameters for the proposed procedure. We utilized the knowledge found in the literature to help determine the contrast value and used the value of 0.5 °C in this case study. Thus, it assumed the temperature contrast less than 0.5 °C in each iteration was contributed to the regional maxima. In addition, when the temperature contrast between delaminated and sound area is less than 0.5 °C, the proposed method could not have enough distinguish power. In terms of this low signal-to-noise ratio, decreasing contrast *h* could increase the distinguishing power, but reduce the reliability of the method. The stopping criterion determined here was assuming no difference between two reconstructed backgrounds larger than the selected contrast *h*. With this assumption, we constrained the processed region within the bridge deck area and excluded the area outside.

## CONCLUSIONS

This paper proposed a pre-processing method to estimate the non-uniform thermal background for delamination detection of the bridge deck. The process was based on grayscale morphologic reconstruction by dilation that the thermal feature of delamination shown as the regional maxima was obtained through subtraction of background from the raw thermal image. Through comparing to the hammer sounding result, the performance of detectability was improved by applying two type of quantitative detection method. It found that the proposed pre-processing method showed the well compatibility to the threshold-based and clustering-based methods for delamination detection. Further work will focus on extending and evaluating the method on more cases to establish a more genetic procedure and general parameter tuning.

## ACKNOWLEDGMENT

The investigation received support from the Nebraska Department of Transportation through facilitating data collection and sharing the non-destructive evaluation results.

## AUTHOR CONTRIBUTION STATEMENT

The authors confirm contribution to the paper as follows:

Study conception and design: Zhigang Shen, Chongsheng Cheng; data collection: Chongsheng Cheng, Zhexiong Shang; analysis and interpretation of results: Chongsheng Cheng, Zhigang Shen; draft manuscript preparation: Chongsheng Cheng. All authors reviewed the results and approved the final version of the manuscript.